\documentstyle[11pt,aaspp4,flushrt]{article}

\def\ref{\par\noindent\hangindent=1in\hangafter=1}
\def\etal{et al.}			
\def\H0{$H_0$~= 75 \kms\ Mpc$^{-1}$}

\def\kms{km s$^{-1}$}

\def\vi{$V\!-\!I$}

\def\sigvi{$\sigma_{V\!-\!I}$}
\def\ref{\par\noindent\hangindent 30pt}

\def\farcs{\hbox{$.\!\!^{\prime\prime}$}}

\def\v16{$\Delta V_{1-6}$}

\begin{document}

\clearpage

\thispagestyle{empty}	


\begin{large}
{\hfill To appear in {\it The Astronomical Journal}, Vol.\ 124, July 2002}
\end{large}

\bigskip
\bigskip
\bigskip
\bigskip
\bigskip

\begin{center}

\begin{Large}

{\bf THE LUMINOSITY FUNCTIONS OF OLD AND INTERMEDIATE-AGE GLOBULAR
CLUSTERS IN NGC 3610$^1$} \\

\end{Large}

\end{center}

\begin{large}			

\bigskip
\bigskip
\bigskip

\centerline{BRADLEY C. WHITMORE}

\centerline{Space Telescope Science Institute, 3700 San Martin Drive,
Baltimore, MD 21218}
\centerline{Electronic mail: whitmore@stsci.edu}

\bigskip
\bigskip

\centerline{FRAN\c COIS SCHWEIZER,}

\centerline{Carnegie Observatories, 813 Santa Barbara St., Pasadena, CA 91101-1292}
\centerline{Electronic mail: schweizer@ociw.edu}

\bigskip
\bigskip

\centerline{ARUNAV KUNDU,}

\centerline{Michigan State University, Physics \&
Astronomy Department, East Lansing, MI 48824-1116}
\centerline{Electronic mail: akundu@pa.msu.edu}

\bigskip
\bigskip

\centerline{and BRYAN W. MILLER}

\centerline{AURA/Gemini Observatory, Casilla 603, La Serena, Chile}
\centerline{Electronic mail: bmiller@gemini.edu}

\bigskip
\bigskip
\bigskip

Received:

\bigskip
\bigskip
\bigskip
\bigskip
\bigskip
\bigskip

\noindent $^{1}$ Based on observations with the NASA/ESA {\it Hubble Space 
Telescope}, obtained at the Space Telescope Science Institute, which is
operated by the Association of Universities for Research in Astronomy,
Inc., under NASA contract NAS5-26555.

\clearpage

\centerline{\bf ABSTRACT}

The Wide Field and Planetary Camera 2 on board the {\it Hubble Space
Telescope} has been used to obtain high-resolution images of NGC 3610,
a dynamically young elliptical galaxy in a group environment. These
observations supersede shorter, undithered HST observations where an
intermediate-age population of globular clusters was first
discovered. The new observations show the bimodal color distribution
of globular clusters more clearly, with peaks at \vi\/ = 0.95 and
1.17. The luminosity function of the blue, metal-poor population of
clusters in NGC 3610 turns over, consistent with a Gaussian
distribution with a peak $M_V$ $\approx$ --7.0, similar to old
globular-cluster populations in elliptical galaxies. The red,
metal-rich population of clusters has a luminosity function which is
more extended toward both the bright and faint ends, as expected for a
cluster population of intermediate age. It is well fit by a power law
$\phi(L)dL \propto L^{\alpha} dL$, with an exponent of $\alpha = -1.78
\pm 0.05$, or $\alpha = -1.90 \pm 0.07$ when corrected for
observational scatter.  A Kolmogorov-Smirnov test confirms the
significant difference between the luminosity functions of the red and
blue clusters, with a probability of less than 0.1 \% that they come
from the same population.  A comparison with the Fall \& Zhang
cluster disruption models shows marginal agreement with the observed
data when comparing both the luminosity functions and the mean color
distributions, although there are differences in detail.  In
particular, there is no clear evidence of the predicted turnover at
the faint end, although deeper observations will be required to make a
definitive test. A by-product of the analysis is the demonstration
that, at any given metallicity, the peak of the luminosity function should remain nearly constant
from 1.5 - 12 Gyr, since the effect of the disruption of faint
clusters is almost perfectly balanced by the fading of the
clusters. This may help explain the apparent universality of the peak
of the globular cluster luminosity function.

\bigskip


\clearpage

\centerline{\bf 1. INTRODUCTION}

Until recently, all globular clusters were considered to be among the
oldest objects in the universe, and various theories were
developed to explain their formation some 15 Gyr ago. However, the
discovery of young massive compact clusters in merging and
starbursting galaxies (e.g., Holtzman \etal\/ 1992, 1996; Whitmore \etal\/
1993; Whitmore \& Schweizer 1995; Meurer \etal\ 1995, Schweizer \etal\/ 1996,
Zepf \etal\/ 1999) provides a
means to study the formation of what appear to be young globular
clusters in the {\it local} universe.

Young merger remnants (e.g., $\la$ 1 Gyr) are generally easy to identify,
both because of their disturbed morphologies and the burst of
star formation that accompanies the interaction. Intermediate-age merger
remnants (e.g., 1 -- 5 Gyr) are more difficult to identify since the
main body has had time to relax into a relatively symmetric configuration
and the newly-formed stellar populations have had time to fade and
redden (Schweizer \& Seitzer 1992). In particular, newly formed metal-rich
globular clusters with ages in the range 1 -- 2 Gyr will have roughly
the same \vi\/ colors as old metal-poor globular
clusters (see Whitmore \etal\/ 1997; hereafter Paper~I), making it
difficult to distinguish the two.  Fortunately, the intermediate-age
clusters will still be 1 -- 2 mag brighter than the old clusters.

Recently, a number of studies have identified intermediate-age
globular clusters in several merger remnants. This helps provide the
``missing link'' between young merger remnants and elliptical
galaxies.  For example, in Paper~I we estimated that NGC 3610 suffered an
interaction 4 $\pm$ 2.5 Gyr ago.  The case for NGC 1700 was less
clear, although the data were compatible with an age of $\sim$4 Gyr.
A recent study by Brown \etal\/ (2000) yields an estimated age of 2 -- 5~Gyr 
for NGC 1700 globulars.  
Similar results have been obtained for
NGC 6702, with Georgakakis, Forbes, \& Brodie (2001) estimating an age
of 2 -- 5 Gyr for its red globular clusters. 
Currently, the best example of an early-type galaxy with intermediate-age
globular clusters is probably NGC 1316, where photometric and
spectroscopic observations of its red clusters yield ages of 3 $\pm$ 0.5 Gyr
(Goudfrooij \etal\/ 2001a,b). 

The luminosity functions for clusters in young mergers, such as ``The
Antennae,'' are power laws with indices of $\alpha\approx -2$
(Whitmore \etal\/ 1999b; Whitmore 2001).  This distribution is markedly
different from the peaked Gaussian profiles found for luminosity
functions of old globular clusters (e.g., Figure~3 of Zhang \& Fall
1999). However, various destruction mechanisms---such as 2-body
evaporation, bulge and disk shocking, and stellar
mass loss---should remove the less massive, more diffuse clusters as
the galaxy ages. Hence, this process may eventually yield a peaked
distribution of clusters similar to what is seen for old globular
clusters.  Intermediate-age remnants, such as NGC 3610 (Scorza \&
Bender 1990, Schweizer \& Seitzer 1992), offer the possibility of
observing this process in action.  While the blue, metal-poor globular
clusters should have a distribution similar to old globular clusters
in elliptical galaxies, peaking at $M_V$ $\approx$ $-$7.2 with a
dispersion of $\sigma$ $\approx$ 1.3 (Whitmore 1997), the red,
metal-rich clusters should extend to both brighter and fainter
luminosities and may show a dropoff one or two magnitudes fainter than
is seen for the old globular clusters (Fall \& Zhang 2001, Goudfrooij
\etal\/ 2001b).

In the following we adopt a Hubble Constant of $H_0 = 75$ km s$^{-1}$
Mpc$^{-1}$ and a mean corrected group velocity of 2251 km s$^{-1}$
(Paper I), which places NGC 3610 at a distance of 30.0 Mpc,
corresponding to a distance modulus of 32.39 mag.  At this distance
the projected scale is 1$''$ = 146 pc, and 1 pixel of the Planetary
Camera ($0\farcs04554$) covers 6.63 pc.

\bigskip
\bigskip

\centerline{\bf 2. OBSERVATIONS AND REDUCTIONS}

\bigskip

\centerline{\it 2.1 WFPC2 Observations}

Our previous WFPC2 observations of NGC 3610 consisted of 2 $\times$
500 s exposures in F555W and 2 $\times$ 600 s exposures in F814W. The
exposures were undithered. The new Cycle 7 observations (Proposal ID =
7468), obtained on November 3, 1999, consisted of 4 $\times$ 1200~s
exposures in F555W and 4 $\times$ 1400 s exposures in F814W. In both
passbands we observed at two positions (offset from each other by $\sim$5.5
pixels in the Planetary Camera (hereafter PC)  and $\sim$2.5 pixels in the 
Wide Field Camera (hereafter WF)) to permit subpixel
dithering. The DRIZZLE software developed by Fruchter \& Hook (1998)
was used to combine the images in each passband, with parameters set
to  {\it pixfrac}~=~0.6 and {\it scale}~= 0.5 (i.e., pixel size).  The factor
of $\sim$5 increase in the total exposure time, the use of dithering,
and the use of a new search algorithm allowed us to reach a factor of
$\sim$2 fainter in luminosity than with our 1994 observations, as can
be seen by comparing the new completeness functions of Figure~1 with
the corresponding functions of Figure~7 in Paper~I.

\bigskip

\centerline{\it 2.2 New Search Algorithm}

Our primary goal for the Cycle 7 observations of NGC 3610 was to push
the detection threshold as faint as possible, in order to determine
whether the luminosity function for the intermediate-age clusters
continues as a power law to fainter magnitudes or turns over, as
predicted by theory (e.g., Fall \& Zhang 2001).
We therefore developed a new search algorithm in order to reach
magnitudes as faint as possible. The first step of this algorithm
uses the DAOFIND task from the daophot package (Stetson 1987) to
identify potential sources down to a very faint limit (2 $\sigma$ of
background) on the $V$, $I$, and summed $V+I$ images.
This results in a relatively large number of false detections in the
bright inner regions of the galaxy. Such detections are then filtered
out by requiring a $\geq$5~$\sigma$ detection on a median-subtracted $V+I$
image (with aperture radii for
object/inner background/outer background set to 2/2/4 pixels on the
dithered image, corresponding to 1/1/2 pixels on the original image).
The algorithm then performs photometry of the surviving candidate
sources on both the drizzled $V$ and $I$ images. Sources are further selected
by requiring a $\ge$5~$\sigma$ detection in one of the two bands, and
values for the concentration index  (the difference in
magnitude using apertures with 1 and 6 drizzled-pixel radii on the PC, or 1 and 4 drizzled pixels on the WF) 
between 1.4 and 2.05 on the WF images
and greater than 1.8 on the PC image. This choice of concentration
indices filters out both hot pixels and background galaxies
(see
Miller et al.\ 1997 for an example of how the size criteria work
for undithered images). Note
that there is no upper limit for the concentration index on the PC
because objects very near the center have large apparent
values of the index, due to the very strong galaxy background.
A visual inspection and simple statistics suggest that most
or all of these objects are globular clusters rather than background
galaxies. 

Figure~2 shows a median-subtracted F555W image of the central region of NGC 
3610.  The apparent warp of the central disk first described in Paper~I is
prominent. The red clusters with \vi~= 1.025 -- 1.5
are circled, while the blue clusters with \vi~=
0.5 -- 1.025 are marked by squares. The bright clusters from Table~1 are
marked with larger symbols and labeled. 
A comparison with our earlier HST data shows that 22 of the 23
clusters in the overlap region with Figure~3b of Paper~I are recovered.
In addition, 11 new candidate clusters are found, primarily in the
inner region where our new search technique makes it possible to identify 
clusters in regions of higher background. These newly discovered objects
include nine of the 50 brightest clusters (i.e., \#1, 4, 8, 14, 15, 16,
29, 47, and 48). Eight of these nine clusters are red, emphasizing the
preponderance of bright red clusters near the center of the galaxy. 

\bigskip

\centerline{\it 2.3 Photometry}

Aperture corrections were determined by measuring magnitudes
through various apertures for several well exposed clusters on each of
the four chips.  The aperture corrections for the PC were $-$0.304 mag
in V and $-$0.363 mag in I, using aperture radii for object/inner
background/outer background of 6/10/16 pixels on the dithered
image (i.e., 3/5/8 pixels on the original image). The corrections on
the WF were $-$0.312 mag in V and $-$0.370 mag in I, using aperture radii
of 4/10/16 pixels on the
dithered image. Small chip-to-chip corrections were also applied using
the values in Table 28.1 of the HST Data Handbook (Voit 1997).

A comparison of our new photometry with the $V$ magnitudes of 40
bright clusters found earlier (Paper~I, Table 4) shows a median
difference of 0.10 mag, with Paper~I values being fainter. The
comparison is especially poor for four clusters very near the center
of the galaxy, where differences in $V$ can be as large as 1
mag. Although we realized in Paper~I that these clusters had large
uncertainties (i.e., error estimates ranging from 0.44 to 1.05 mag),
the main reason for this difference now appears to be our use of
larger apertures for both the objects and the background in the older
observations. The very strong background gradients in the inner
regions and the larger sky annuli resulted in our overestimating the
background in the Paper~I photometry.  After removing the outliers
near the center of the galaxy the dispersion of the differences in $V$
is 0.09 mag, with a mean difference 0.07 (Paper I values fainter).  The comparison
for \vi\ colors is better, with a mean difference of
only 0.03 mag (Paper I values redder) and a dispersion of 0.09 mag.

Completeness tests were performed for five background levels on the PC
and four background levels on the WFC by using the task {\it addstar} in
the DAOPHOT package. This program adds artificial objects derived by
using high S/N images of isolated clusters from the same chip. Figure~1
shows the resulting completeness functions. The completeness thresholds used
in the present paper are defined as the magnitude at which only half the
artificial objects are identified.  Corrections for non-optimal
charge transfer efficiency on the CCDs of WFPC2 were applied using the
formulae by Whitmore, Heyer, \& Casertano (1999). Finally, the
instrumental magnitudes were transformed to the Kron-Cousins system
via the formulae by Holtzman \etal\/ (1996).

\bigskip
\bigskip

\centerline{\bf 3. ANALYSIS}

\bigskip

\centerline{\it 3.1 Color Distribution}

Figure~3 shows the new \vi\/ vs. $V$ color-magnitude diagram for NGC
3610 clusters. This diagram looks quite similar to the corresponding
diagram of Paper~I, but now extends to a fainter detection limit of $V
\approx 26.5$ at 50\% completeness . We again find a clear trend for
the brighter clusters to be redder.

Figure~4 shows histograms of \vi\/ indices of candidate clusters
within 50$''$ from the center of NGC 3610.  The {\it left panel} shows
all candidate clusters brighter than $V=26.5$ while the {\it right
panel} shows only those brighter than $V=25$.  Note that the clusters
in the PC (i.e., the area shaded black) show clearly bimodal color
distributions in either case, while in the combined PC\,+\,WF area the
brighter clusters show bimodality more cleary, presumably
because of the higher signal-to-noise.  This diagram was
obtained as follows.  All candidate clusters within 50$''$ from the
center and with \sigvi~$\leq 0.4$~mag were binned in 0.05-mag
intervals and corrected for contamination by background objects.  The
latter correction was derived by selecting objects beyond 85$''$ from
the center according to the same selection criteria, smoothing their
color distribution with a 7-bin triangular weight function to diminish
small-number noise, and subtracting their surface-number density from
that of the clusters within the central 50$''$.  The corrected numbers
of clusters is shown as a solid line in Figure~4, while the numbers
before background correction are indicated by a dotted line.  Since
the background-object selection region beyond 85$''$ presumably still
contains some halo globular clusters, the correction for contamination
by background objects is a maximum, and the true color distribution of
globulars in NGC 3610 must lie somewhere between the solid and dotted
lines (likely close to the solid lines).

The most central dozen
of the selected clusters all lie close to, or within the, central
warped stellar disk (Figure~2). Although all but two of these twelve
clusters look round and symmetric, one might question whether some of
them are disk clumps or noise of uniform color and, therefore,
contribute unduly to the red peak of the bimodal color distribution.
To check for this possibility, we excluded these twelve clusters by
requiring that the level of the surrounding background in $V$ be less
than 130 counts per pixel.  The resulting color distributions looks
nearly identical to Figure~4.  We conclude that the very pronounced red peak
of the central clusters is not an artifact of disk noise, but simply
reflects the strong central concentration of this subpopulation of
globulars.

The bimodality is more pronounced
than in Paper 1, perhaps because of smaller measuring errors
due to the longer exposures and dithered data. We used
the KMM test developed by Ashman, Bird, \& Zepf (1994) to objectively
estimate the probability of a bimodal distribution. A wide range of
selection criteria were employed to test the robustness of the
estimates.  In essentially all cases the distribution was found to
be bimodal, with probabilities ranging between 93\% and 97\% if we restrict
the sample to 0.7 $<$ \vi\/ $<$ 1.5, $\sigma(V-I) < 0.3$, and use
projected distances from the center of less than 20$''$ (n=39), 30$''$
(n=57), and 40$''$ (n=74). The mean values for the two peaks of the
distributions fall in the ranges \vi\ = 0.94 -- 0.96 and
1.16 -- 1.17 for these three spatial subsamples,
very similar to the values found for many elliptical galaxies (e.g.,
Kundu \& Whitmore 2001).  Roughly 30\% of the clusters appear to be
members of the bluer, metal-poor population, according to the
KMM test.  If we extend the sample to larger projected distances the
bimodality probabilities
become smaller (e.g., 82\% for $R <$ 50$''$ and 74\% for the
full field of view), presumably because of increasing contamination
by background objects in the outer regions, and perhaps also
because of larger photometric errors for the fainter
clusters, which tend to be found in the outer regions.

A close look at Figure~3 shows that most of the clusters in the
brightest 2 -- 3 mag range belong to the red, metal-rich
population, with colors in the range \vi~= 1.025 -- 1.30. However, at a
magnitude of  V $\approx $ 25 it appears that the dominant
population is the blue, metal-poor clusters, with colors in the
range 0.80 -- 1.025. We note that a typical value of $M_V$ for an old
globular cluster population is --7.2 (Whitmore 1997). Hence
we expect a peak in the blue population at $M_V$ $\approx$ 25.2, given
the distance modulus of $m-M = 32.39$.  Near the completeness limit at
$V \approx 26.5$, the metal-rich population again appears to dominate.
This alternating dominance of red and blue clusters is
consistent with expectations from a superposition of an old metal-poor
population of globular clusters and an intermediate-age metal-rich
population of globular clusters, if the latter has a luminosity function
extending farther to both brighter and fainter magnitudes (i.e.,
more similar to a power law).

The statistical significance of this trend can be better quantified by
computing the mean \vi\/ as a function of V, as shown in Figure~5. We
first trim the distribution to only include objects in the range
$0.8 < V-I < 1.3$, since Figures 3 and 4 show that most candidate
clusters fall in this range. Figure 5 {\it (top)} shows the \vi\/ means
without any restriction on the radius. This diagram supports the first
visual impression from the color-magnitude diagram: the brightest clusters
at $V \approx 22$ have a mean \vi\/ = 1.15 $\pm$ 0.04.  The mean then
migrates to \vi\/ = 1.05 $\pm$ 0.02 at $V \approx 25$, and finally shifts
back toward the red with \vi\/ = 1.09 $\pm$ 0.02 at $V$ = 26.25. Excluding
clusters in the range $V = 24$ -- 26, the data represent
a 5 $\sigma$ deviation from a value of \vi\/ = 1.05.

If we restrict the sample to objects within
$r < 20''$ (middle panel in Fig.~5), we once again find the brighter
clusters with redder colors. However, objects at fainter magnitudes are
missing due to the brighter completeness correction limit of the PC.

Finally, if we restrict the sample to objects within the annulus of
$20'' < r < 50''$ (bottom panel in Fig.~5), we again find the
bell-shaped curve indicative of redder mean
colors at both the bright and faint ends. In addition, we are able to
observe to fainter magnitudes since the objects lie on the WF chips.
Here we find a continuation of the trend to redder colors, with mean
\vi\/ = 1.11 $\pm$ 0.02 at $V = 27.2$.

\bigskip\medskip

\centerline{\it 3.2 Luminosity Functions of Blue and Red Clusters}

The derivation of the luminosity function (hereafter LF) of a cluster
population on a strongly varying galaxy background presents a serious
challenge.  The detected-cluster counts need to be corrected for
incompleteness as a function of both the cluster magnitude and the
local level of galaxy background.  Yet, the experimentally determined
completeness functions of Figure~1 (\S 2.3) show a strong dependence on
the background level and differ sharply between the PC and WF cameras.
As a result, piecing together an incompleteness-corrected luminosity
function from clusters imaged by the two different cameras and
lying on backgrounds of very different brightness is nearly impossible.

The problem is alleviated if one restricts oneself to clusters
imaged on the WF chips.  Not only are the background variations much
reduced (since NGC 3610 was centered on the PC), but equal detection
probability is reached at $V$ magnitudes nearly 1.0~mag fainter on
the WF images than on the PC image.  Therefore, we restrict our
discussion of the LF of globular clusters in NGC 3610 to the LF
determined from clusters imaged on Chips WF2\,--\,WF4. The LF on the PC is
consistent with the following results, but the low number statistics
make it impossible to draw any firm conclusions.

Figure~6 shows the luminosity functions of blue and red globular
clusters determined separately.  The LFs corrected for contamination
by background objects are shown as solid lines and are shaded black
above the 50\%-completeness limits (marked by dashed vertical lines).
For comparison, the LFs {\it un}corrected for background contamination
are shown as dotted lines.

These luminosity functions were determined as follows.  First, candidate
clusters were separated into two subsamples of blue ($0.80 \leq
V\!-\!I \leq 1.025$) and red ($1.025 < V\!-\!I \leq 1.30$) clusters.
For each of these subsamples, all candidate clusters imaged on WF chips
and lying within a limiting radius $R_{\rm lim}\!= 50''$, 65$''$, or 80$''$
from the center were selected, subject to the conditions that
$19 \leq V \leq 28$, \sigvi~$< 0.60$~mag, and the galaxy-background level
be $0.1 \leq V_{\rm bkg} \leq 100$ counts/pixel.  Each cluster was
then corrected for incompleteness by dividing its count of one by the
completeness fraction corresponding to its $V$ magnitude and background
level.  This fraction itself was obtained from the functions shown in
Figure~1 by bilinear interpolation. The corrected (i.e.,
weighted) cluster count was then added to the appropriate bin in
$V$ magnitude, with bin sizes of typically 0.25~mag.

The resulting raw luminosity functions were further corrected for
contamination by fore- and background objects (i.e., stars and distant
galaxies).  To that effect, objects were selected at distances
$R_{\rm bkg} > 85''$ from the center and subject to the same color,
magnitude, and galaxy-background selection criteria as the objects in
the inner search zones.
The ``luminosity functions'' of these objects were determined in a
similar manner as described above, then smoothed by a 7-bin triangular weighting
function to diminish small-number noise, scaled to the areas of the inner
search zones, and subtracted from the above raw LFs.  It is these corrected
LFs that are shown as solid lines (and shaded black above the 50\%
completeness limits) in Figure~6.

As Figure~6 illustrates, the luminosity functions of blue and red
globular clusters in NGC 3610 differ dramatically.  While the LF for
the blue globulars is approximately Gaussian, that for the red
globulars is more nearly power-law shaped.  When compared to the blue
clusters, there are more red clusters both at the bright end and near
the 50\%-completeness limit, consistent with the mean colors discussed
in \S 3.1 and shown in Figure~5.  The LFs shown are for 205 candidate
globular clusters (64 blue, 141 red) on the WF chips out to $R_{\rm
lim}\!= 80''$.  The same shape difference is observed for the LFs of
clusters out to $R_{\rm lim}\!= 50''$ and 65$''$, though with more
noise due to smaller cluster numbers.

The statistical significance of this result can be better quantified
by performing a Kolmogorov-Smirnov test on the populations of red
globular clusters and blue globular clusters, as defined above.
Figure~7 shows the results based on the completeness- and
background-corrected luminosity functions from Figure 6.  The top
panel shows the fraction of each distribution as a function of $V$
magnitude for the clusters on the WF chip.  The value of the largest
discrepancy is shown on the figure, as is the probability that the two
populations come from the same distribution (i.e., only a 1.1 \%
chance). We note that the nature of the trend, with more red globular
cluster's at both the bright and faint ends, tends to cancel out the
value of the largest discrepancy. The two lower panels in Figure~7
show the results if we break the cumulative distributions in two at
the cross-over point, $V$ $\approx$ 25. The largest fractional
discrepancies are now $\approx$ 0.4. While the small number of
clusters in the bright part of the WF sample results in a relatively
large value of the probability (2.1 \%), the faint sample now shows a
very small probability (0.09 \%) that the red and blue globular
clusters are drawn from the same sample.

\bigskip\medskip

\centerline{\it 3.3 Comparison with the Fall \& Zhang (2001) Cluster Disruption
Models}

Fall and Zhang (2001) have developed models for the disruption of
globular clusters, which include two-body relaxation, gravitational
shocks, and stellar evolution.   Figure~4 in Fall \&
Zhang (2001) demonstrates how an initial power-law or Schechter  mass function will
evolve toward a peaked distribution, typical of an old globular
cluster population. As the fainter clusters are destroyed, the peak of
the distribution shifts toward larger mass. However, the luminosities
of the clusters fade at almost exactly the same rate (e.g., see Figure~13
in Paper I), hence compensating for the effect. The net result is that, at a given
metallicity,
the peak luminosity remains almost constant with time in the range 
1.5 -- 12 Gyr. This may help explain the apparent universaility of the 
globular cluster luminosity function, as demonstrated
by various studies (e.g., Harris, 1991)

Figure~8 demonstrates this effect.  We combined the Fall \& Zhang
(2001) model that employs a Schechter initial mass function (i.e., the
bottom left panel in their Figure~4) and the prediction for fading
from the Bruzual \& Charlot (1996, unpublished) stellar evolution models in order
to predict the relative shifts between the peaks of the various
distributions. Solar metallicity is used for the young and
intermediate-age clusters while 0.02 solar metallicity is used for the
old (12 Gyr) population. The models are normalized at $V$ = 25. The
top panel shows the luminosity function on a log scale while the
bottom panel shows it on a linear scale. The predicted peaks for the
1.5, 3, and 6 Gyr models are essentially identical.  The only reason
for the shift in the peak of the 12 Gyr old population is the lower
metallicity. If we were to use solar rather than 0.02 solar
metallicity for the 12 Gyr population, we would find that the peaks of
the distributions vary by only $\sim$ 0.10 mag over the full range from
1.5 -- 12 Gyr.

The tail of the predicted LF at the faint end is nearly identical for
all three intermediate ages, but the bright end shows some
variation. In particular, the 1.5 Gyr population is predicted to have
more bright clusters than the 12 Gyr old clusters while the 6 Gyr
population is predicted to have fewer. Based on this fact alone it
would appear that the intermediate-age clusters in NGC 3610 would fall
in the range 1.5 -- 3 Gyr. We note that the turnover in the 12 Gyr
population is predicted to be about 1.5 magnitudes brighter than our
completeness limit of $\sim$26.6, but the predicted turnover in the
intermediate-age population is only $\approx$ 0.5 magnitudes
brighter. Figure 6 does not show any clear evidence for a turnover
of the red population out to the completeness limit, although there is
a hint that it may flatten out beyond this point. Deeper observations
will be required to make a definitive test, however.

While the young clusters in recent mergers  appear to have roughly
solar metallicity (e.g., Schweizer \& Seitzer, 1998), the metallicity
for older remnants is more uncertain, and may tend to be lower since
the merger event happened further in the past. 
If we adopt 0.4 solar for the metallicity of the 1.5 -- 6 Gyr populations
we would find that the peak luminosities move toward brighter
magnitudes by $\sim$ 0.3 mag, reducing only slightly the difference between the
metal-poor 12 Gyr population and the metal-rich younger populations.

These models can also be used to predict how
the values of mean (\vi\/) change with magnitude.
Figure~9 shows the predicted mean colors, based on the models in
Figure~8. We find the same qualitative behavior as shown by our data
for NGC 3610 in Figure~5, with redder colors for both bright and faint
clusters and a mean color $\approx$ 1.05 in the intermediate
range. The best agreement appears to be with the 1.5 Gyr
models. However, we note that the bluest mean colors are predicted to
be in the range $V \approx$ 23 -- 24, while the observed peak is
$V \approx 25.4$. 
One possible explanation for this difference might be the fact that the models assume a  zero-age burst
of cluster formation, while there is good evidence --- both observationally
(Whitmore et al. 1999b) and theoretically (Mihos et al. 1993) --- that
clusters can be produced for roughly half a billion years
during a merger.

\bigskip\medskip

\centerline{\it 3.4 Fits and Simulations of the Blue and Red  Populations}

Figure 10 shows a Gaussian fit to the blue clusters (upper panel) and
a variety of power-law and Fall \& Zhang (2001) model fits to the red
clusters (lower panel).  The Gaussian fit to the blue population
yields a peak at $V = 25.44 \pm 0.10$ and a width $\sigma$ = 0.66
$\pm$ 0.19.  This corresponds to a value of M$_V = -6.95$, similar to
the value found for normal elliptical galaxies (e.g., Whitmore 1997
finds M$_V = -7.21 \pm 0.26$).  We note, however, that the value of
$\sigma$ is considerably smaller than typically found for giant
elliptical galaxies (e.g., 1.1 -- 1.5, according to Kundu \& Whitmore
2001).  Fixing $\sigma$ to these values yields peaks in a range from
$V = 25.61 \pm 0.15$ (using $\sigma = 1.1$) to $V = 26.04 \pm 0.3$
(using $\sigma = 1.5$).  However, as Figure 10 shows, even the fit
with $\sigma = 1.1$ is clearly too wide, hence we adopt the original
value of $V = 25.44$ as our best-fit.

The lower panel of Figure 10 includes a best-fit power law $\phi(L)dL
\propto L^{\alpha} dL$, with an exponent of $\alpha = -1.78 \pm 0.05$ (thick solid line),
and a set of Fall \& Zhang (2001) models ranging from zero age to 12
Gyr. The best fitting Fall \& Zhang model is the 1.5 Gyr model (dotted
line). The fits to the 0 and 3 Gyr models
are quite similar while the fits to the 6 and 12 Gyr population are
clearly unreasonable.  Hence, the simple power law provides the best
overall fit to the red population, partly because the bright end
slopes of the Fall \& Zhang models using the Schechter initial mass
distribution appear to be a bit too steep.

Fall \& Zhang (2001) also calculated models using a shallower
power-law of index --2 for the initial mass distribution (i.e., the top left
diagram in their Figure 4). Figure 11 shows the fits to the data using
these models. While the agreement is better at the bright end, the fit
on the faint end is essentially unchanged.  In particular, there is no
clear evidence for the predicted flattening.  However, the current
data is not deep enough to make a definitive test.

Goudfrooij \etal\/ (2001a) claimed that the luminosity function for
the red clusters in the 3 Gyr old merger remnant NGC 1316 has a
shallower slope, with $\alpha = -1.2~\pm~0.3$. However, a comparison
of their Figure 13a with our Figure 6 shows that the two luminosity
functions are actually quite similar. It appears that an error was
made in the calculation of $\alpha$. Goudfrooij (2002; private
communication) has corrected the problem and now finds $\alpha \approx
-1.7~\pm~0.1$ for the red clusters in NGC 1316, which is similar to
the value we find in NGC 3610. We note that the fits are actually
computed by using the log Number vs. $V$ correlation (e.g., the bottom
plot in Figure 10) and then converting it to the form
$\phi(L)dL \propto L^{\alpha} dL$ via the formula $\alpha = 2.5 \times
\beta +1$, where $\beta$ is the slope in the log Number vs. $V$
diagram.

One of the difficulties of extracting the luminosity function for the
red and blue clusters is the necessity of defining color bins which
are smaller than the observational uncertainies at faint
magnitudes. For example, this means that an intrinsically blue cluster
may fall outside the 0.80 $< V < $ 1.025 range simply due to large
observational measuring errors. Conversely, an intrinsically red cluster
may be counted as a blue cluster because it is scattered
into the blue bin.

We can correct for this effect by using simulations similar to those
developed in Paper I. These are designed to approximately match
the color-magnitude diagram, the total luminosity function, and the
color histogram for the brighter clusters. We then use these
simulations to determine correction fractions for the blue and 
red populations as a function of magnitude.

For the blue clusters we adopt a
Gaussian mass function with parameters designed to approximately match
the fit shown in the upper panel of Figure 10, an age of 15 Gyr, and a
metallicity [Fe/H] = --1.3. This results in a mean value of $V-I$ =
0.95.  For the red clusters we adopt a power-law mass function with
$\alpha = -1.8$, an age of 6 Gyr, and metallicity [Fe/H] = 0.0. This
results in a mean value of $V-I$ = 1.17.  We scale the observational
uncertainies so that they mimic the scatter in the color-magnitude
diagram and color histogram.  The resulting age determination of 6 Gyr
is consistent with our estimate of 4 $\pm$ 2.5 Gyr from Paper I, but is
larger than the estimate of 1.5 Gyr based on a comparison with the
Fall \& Zhang (2001) models. However, we note that the uncertainties
in both estimates are roughly a factor of two, hence there is no
serious discrepancy in the comparison.

Figure 12 shows the color magnitude diagram for the simulation. This
demonstrates how the data points for some blue and red clusters are
scattered into regions outside the appropriate color bins. We note
that Figure 12 looks somewhat different than Figure 3 for two
reasons. First, the open and filled symbols in Figure 12 are the
intrinsic blue and red clusters while in Figure 3 they are the
clusters on the PC and WF. Second, Figure 3 includes background
objects that are not included in the simulation.

Figure 13 shows both the intrinsic (dotted line) and apparent (solid
line) luminosity functions for the simulation from Figure 12. This
demonstrates how observational scatter can affect our determinations
of the LFs for the blue and red populations.  In particular, the
scatter  results in an underestimate in the slope of the power law
for the red population, although the effect is relatively
minor (i.e., the value of $\alpha$ changes from 1.77 $\pm$ 0.06
to 1.90 $\pm$ 0.07).

\bigskip
\bigskip

\centerline{\bf 4. SUMMARY}

New, deeper, dithered observations using the WFPC2 on board the {\it
Hubble Space Telescope} have been used to extend our study of NGC
3610, an elliptical galaxy that is a likely intermediate-age merger remnant. Our primary results are the
following.

1. The inner globular clusters show a clear bimodal color
distribution, with peaks at \vi\/ = 0.95 and 1.17. The KMM test
developed by Ashman, Bird, \& Zepf (1994) yields probabilities for
bimodality ranging between 93\% and 97\% if we restrict the
sample to have  projected
distances from the center of less than 20$''$ (n=39), 30$''$ (n=57),
and 40$''$ (n=74). Bimodality probabilities become smaller in the
outer regions (e.g., 74\% for the full field of view), presumably
because of increasing contamination by background objects and larger
photometric errors for fainter objects.

2. The luminosity function of the blue, metal-poor population (i.e.,
0.8 $<$ \vi\/ $\leq$ 1.025) of clusters in NGC 3610 turns over,
with a peak $M_V \approx -7$, similar
to values typical  for old
globular-cluster populations in elliptical galaxies.

3. The red, metal-rich population (1.025 $\leq$ \vi\/ $<$ 1.3) of
clusters has a luminosity function which is more extended toward both
the bright and faint ends, as expected for an intermediate-age cluster
population.  This is evident in both the luminosity function and the
distribution of mean colors.  The red population is well fit by a
power law $\phi(L)dL \propto L^{\alpha} dL$, with an exponent of
$\alpha = -1.78 \pm 0.05$, or
$\alpha = -1.90 \pm 0.07$ when corrected for observational scatter
using a simulation of the data. A Kolmogorov-Smirnov test indicates that
the probability that the red and blue clusters come from the same
population in NGC~3610 is less than 0.1 \% .

4. A comparison with the Fall \& Zhang (2001) cluster disruption
models shows marginal agreement with the observed data when
comparing both the luminosity functions and the mean color
distributions. For example, the model predicts the bluest mean colors
should occur in the range $V \approx$ 23 -- 24 while the observed peak
occurs at $V$ $\approx$ 25.4. The models also predict a flattening and
turnover of the peak in the LF of the red clusters at $V$ $\approx$ 26.  
While there is no clear evidence
of the predicted turnover, deeper observations will
be required to make a definitive test.

5. A by-product of the analysis is the demonstration
that, at any given metallicity, the peak of the luminosity function should remain nearly constant
from 1.5 - 12 Gyr, since the effect of the disruption of faint clusters
is almost perfectly balanced by the fading of the clusters. This may
help explain the apparent universality of the peak of the 
globular cluster luminosity function.

\bigskip
\bigskip

We thank Paul Goudfrooij and Michael Fall for useful discussions.
This work was supported by NASA grant GO-07468.01-A.  Partial support
by the NSF through Grant AST~99-00742 to F.S. is also gratefully
acknowledged. Partial support to B.W.M. was provided by the
Gemini Observatory, which is operated by the Association of
Universities for Research in Astronomy, Inc., on behalf of the
international Gemini parthership of Argentina, Australia, Brazil,
Canada, Chile, the United Kingdom and the United States of America.
AK acknowledges support from NASA LTSA grant
NAG5-11319, and from NASA grant HST-AR-08755.04-A.

\bigskip
\bigskip

\clearpage

\parskip 6pt plus 3pt minus 2pt



\clearpage 

\centerline{\bf FIGURE CAPTIONS}

\bigskip

\noindent Figure 1 - Completeness curves as determined from artificial
star experiments for different background levels on the PC and WF.
The background levels are given in units of DN (Data Number).

\bigskip

\noindent Figure 2 - Median-subtracted image in F555W showing the inner
region of NGC 3610. Squares are blue clusters while circles are
red clusters. Larger symbols are used for the 50 brightest
clusters.

\bigskip

\noindent Figure 3 - \vi\/ vs. $V$ color-magnitude diagram for cluster
candidates in NGC 3610. Open circles are used for the objects on the
PC while filled circles are used for the WF.

\bigskip

\noindent Figure 4 - 
Color distributions of NGC 3610 candidate clusters within 50$''$
from the center; all clusters (solid lines) and clusters on PC chip only
(shaded black). The dotted lines show clusters before correction for
background-object contamination.  {\it Left panel} shows all candidate
clusters brighter than $V=26.5$, while {\it right panel} shows only
those brighter than $V=25$.  Note that the clusters in the PC area
show clearly bimodal color distributions in either case, while
in the combined PC\,+\,WF area the brighter clusters show bimodality
more cleary.

\bigskip

\noindent Figure 5 - Mean values of \vi\/ for clusters in the range
0.8 $<$ \vi\/ $<$ 1.3 as a function of $V$ for various radial ranges
in NGC3610. In all cases the smallest mean value is near $V$ = 25,
with higher values on both the bright and faint ends. This shows that
the distribution of red globular clusters is more extended (i.e., is
more like a power law) than the distribution of blue globular clusters
(which is more like a Gaussian).

\bigskip

\noindent Figure 6 - Luminosity functions of blue ({\it left}) and red
({\it right}) globular clusters in NGC 3610 within 80$''$ from the center.  
The 50\% completeness
limits ({\it vertical dashed}) mark the boundary up to which the LFs can
be trusted.  Areas {\it shaded dark\,} mark background-corrected LFs,
{\it dotted lines\,} the uncorrected LFs.  Notice the nearly lognormal
shape of LF for blue globulars and the nearly power-law LF for red
globulars.  

\bigskip

\noindent Figure 7 - Kolmogorov-Smirnov tests comparing the
populations of blue globular cluster (0.8 $<$ \vi\/ $\leq$ 1.025) and
red globular cluster (1.025 $<$ \vi\/ $\leq$ 1.3). The largest
discrepancies are listed along with the probability that the two
samples come from the same population. The cross-over near $V$ = 25 tends
to cancel out the differences in the two populations, hence the bottom
two panels break the cumulative distribution in half. In all cases we
find strong evidence that the two populations are statistically
different.

\bigskip

\noindent Figure 8 - Predicted luminosity functions (log in the top
panel; linear in the bottom panel) based on a combination of the Fall
\& Zhang (2001) cluster disruption models and the Bruzual \& Charlot
(1996) population synthesis models. Solar metallicity is assumed for
the young and intermediate-age clusters while 0.02 solar is assumed
for the 12 Gyr population. Note that the fading predicted by the BC96
models almost exactly cancels out the increase in the mean mass
predicted by the Fall and Zhang models, resulting in nearly identical
distributions at the faint end for the 1.5, 3, and 6 Gyr models.

\bigskip

\noindent Figure 9 - Predicted values of mean \vi\/ for the models
shown in Figure~8. Note that while the 1.5 and 3 Gyr models have the
same qualitative behavior as shown in Figure~5, the predicted position
of the peak is bluer than found for the observations of NGC 3610.

\bigskip

\noindent Figure 10 - Fits to the luminosity functions of blue ({\it
upper}) and red ({\it lower}) globular clusters in NGC 3610.  The
upper panel shows our best-fit Gaussian (thick solid line; $V = 25.44
\pm 0.10$, $\sigma$ = 0.66 $\pm$ 0.19) for the blue clusters, along
with a fit with $\sigma$ = 1.1 (dotted line), which is clearly
unreasonable.  The lower panel shows a best-fit power law (thick solid
line; slope = --1.78 $\pm$ 0.05), and a range of Fall \& Zhang (2001)
models using Schechter initial mass functions and ages ranging from
zero age (thin solid line) to 12 Gyr (3dot - dash line). The best
fitting Fall \& Zhang model is the 1.5 Gyr model (dotted line).  The 0
and 3 Gyr models give similar results, while the 6 and 12 Gyr models
result in unreasonable fits.

\bigskip

\noindent Figure 11 - Same as Figure 10, but using Zhang \& Fall
(2001) models with a power-law distribution for the
initial mass function.

\bigskip

\noindent Figure 12 - A simulation of the 
\vi\/ vs. $V$ color-magnitude diagram for cluster
candidates in NGC 3610. Filled circles are intrinsically
red clusters while open circles are intrinsically
blue clusters. Note how observational measuring
errors can scatter the data points into the wrong color
bins. See text for details about the simulation parameters.

\bigskip

\noindent Figure 13 - The intrinsic (dotted line) and apparent (solid line)
luminosity functions for the simulation shown in Figure 12. 
This shows that observational scatter can result in an overestimate
in the value of $V$ for the peak of the blue population and
an underestimate in the slope of the power law for the red population.

\clearpage 

\begin{figure*}
\includegraphics{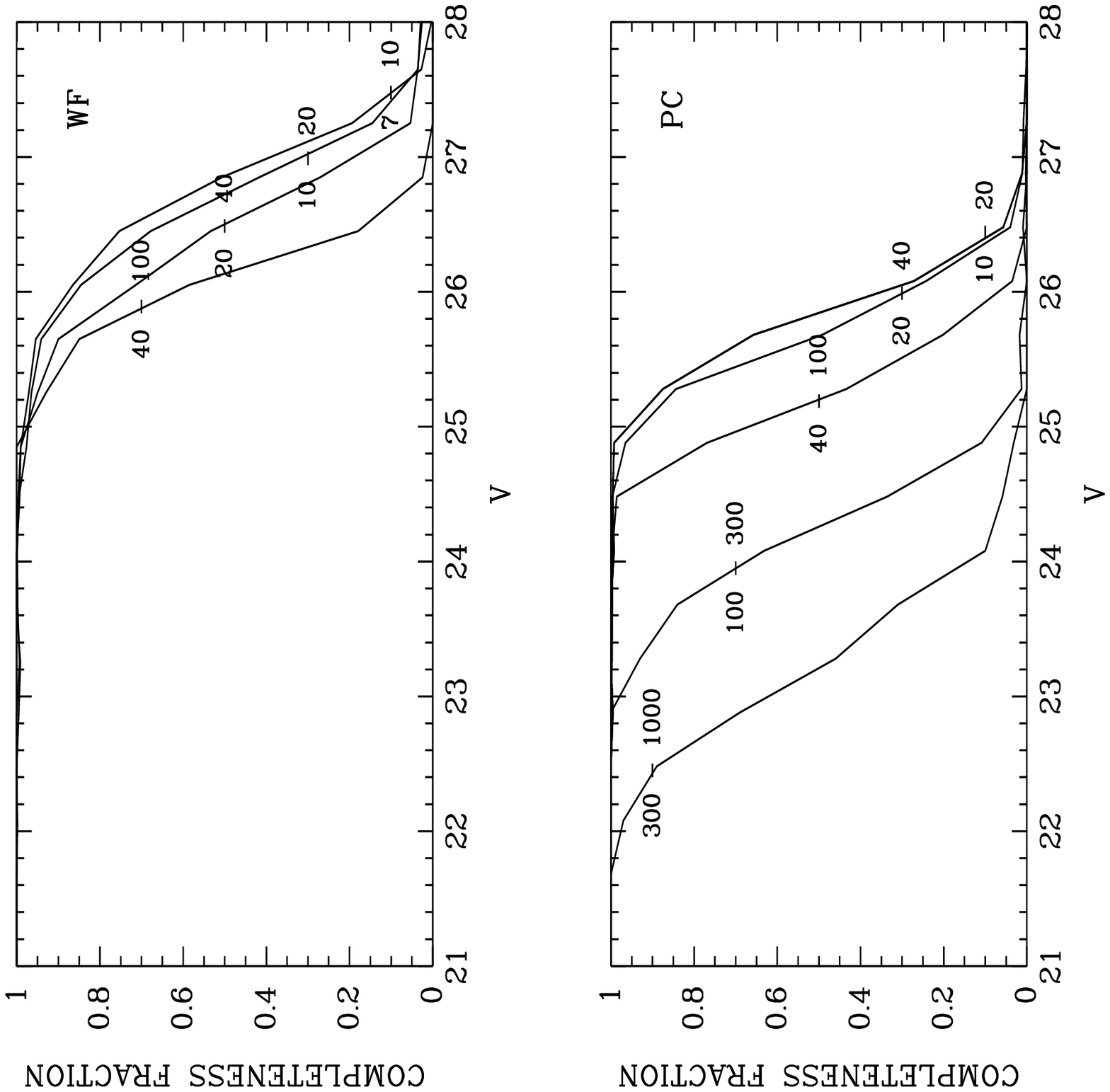}
\vspace{17.0cm}
\caption{}
\end{figure*}

\clearpage 
\begin{figure*}
\includegraphics{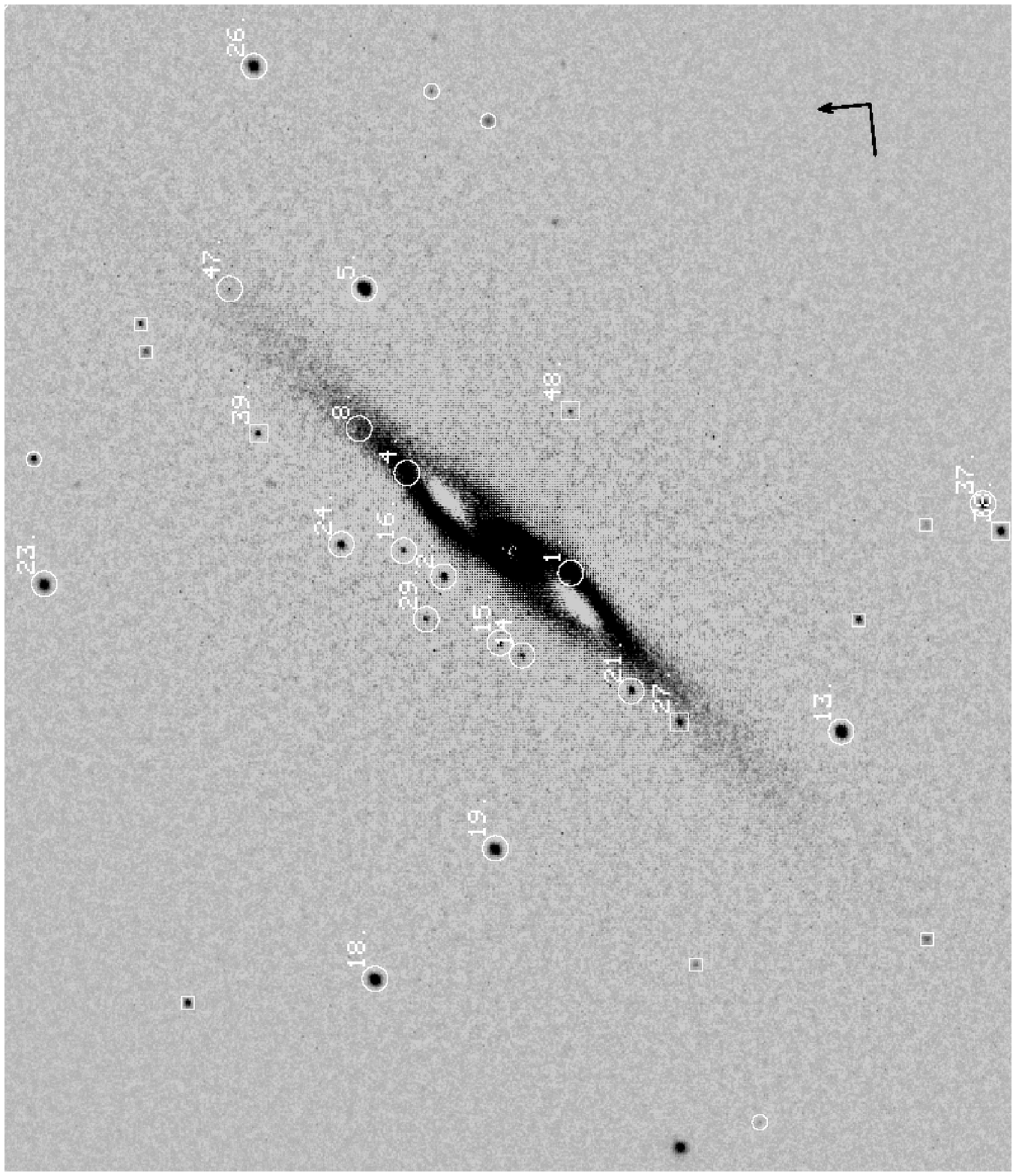}
\vspace{17.0cm}
\caption{ }
\end{figure*}

\clearpage
\begin{figure*}
\includegraphics{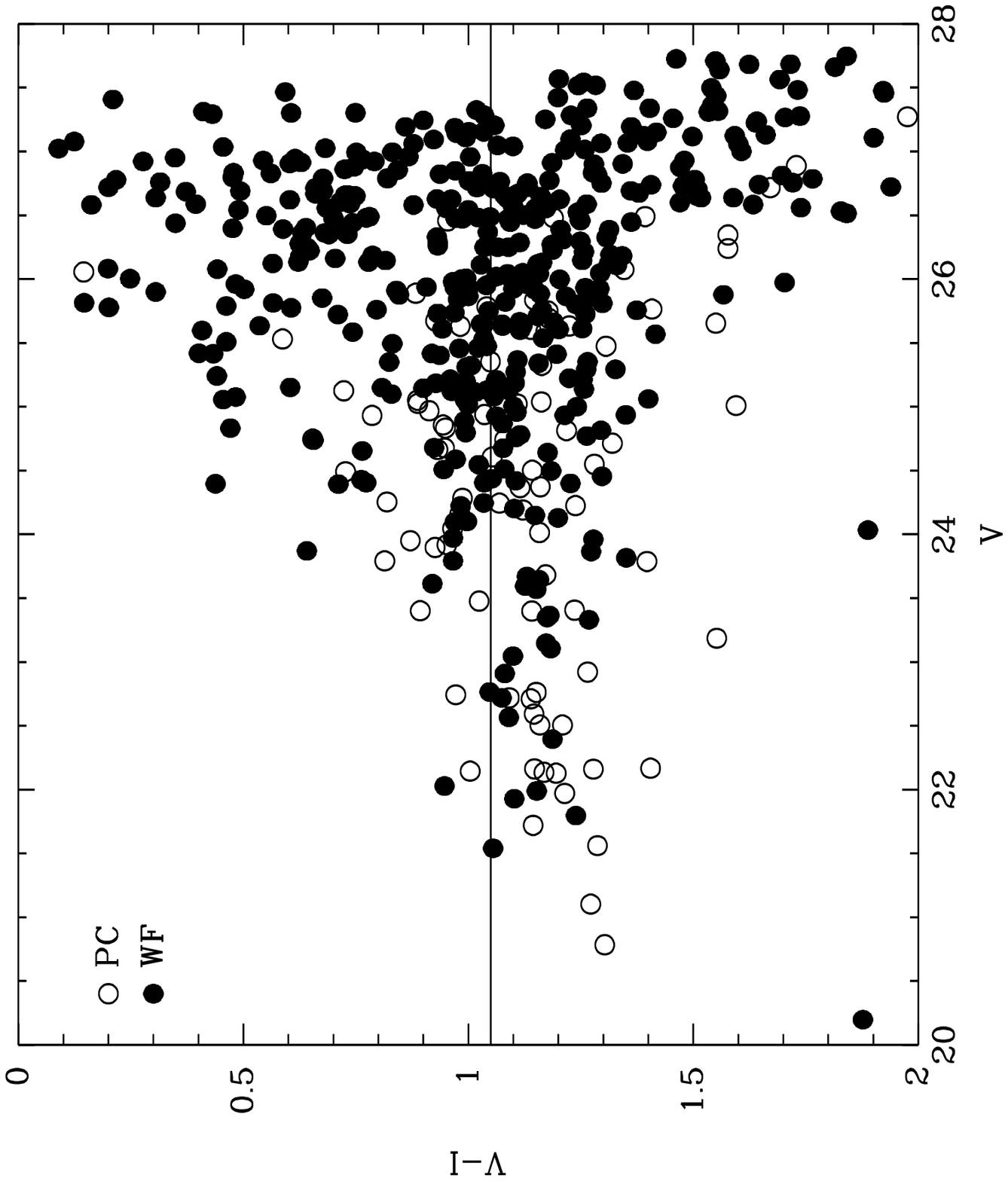}
\vspace{17.0cm}
\caption{}
\end{figure*}

\clearpage
\begin{figure*}
\includegraphics{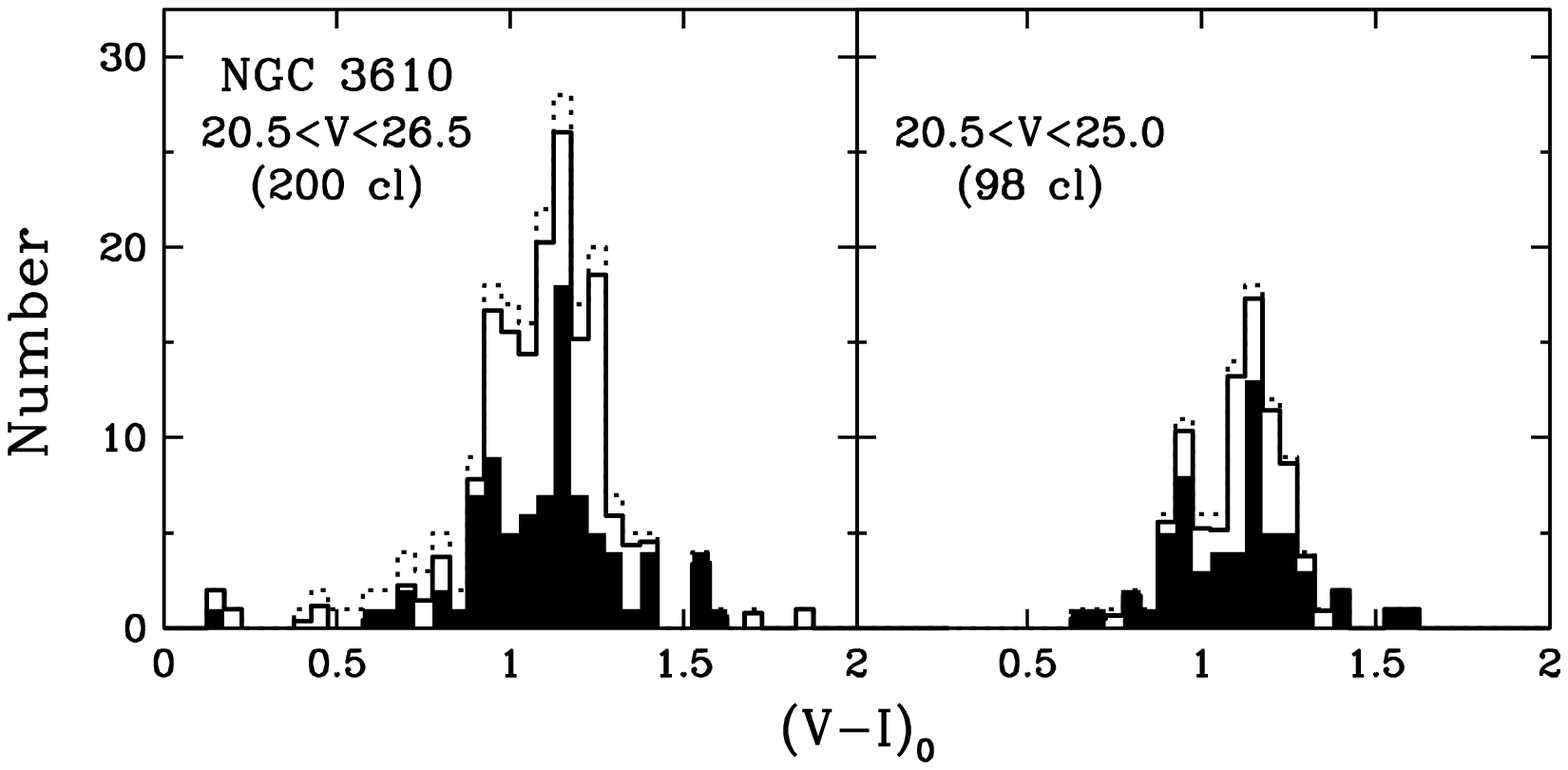}
\vspace{17.0cm}
\caption{}
\end{figure*}

\clearpage
\begin{figure*}
\includegraphics{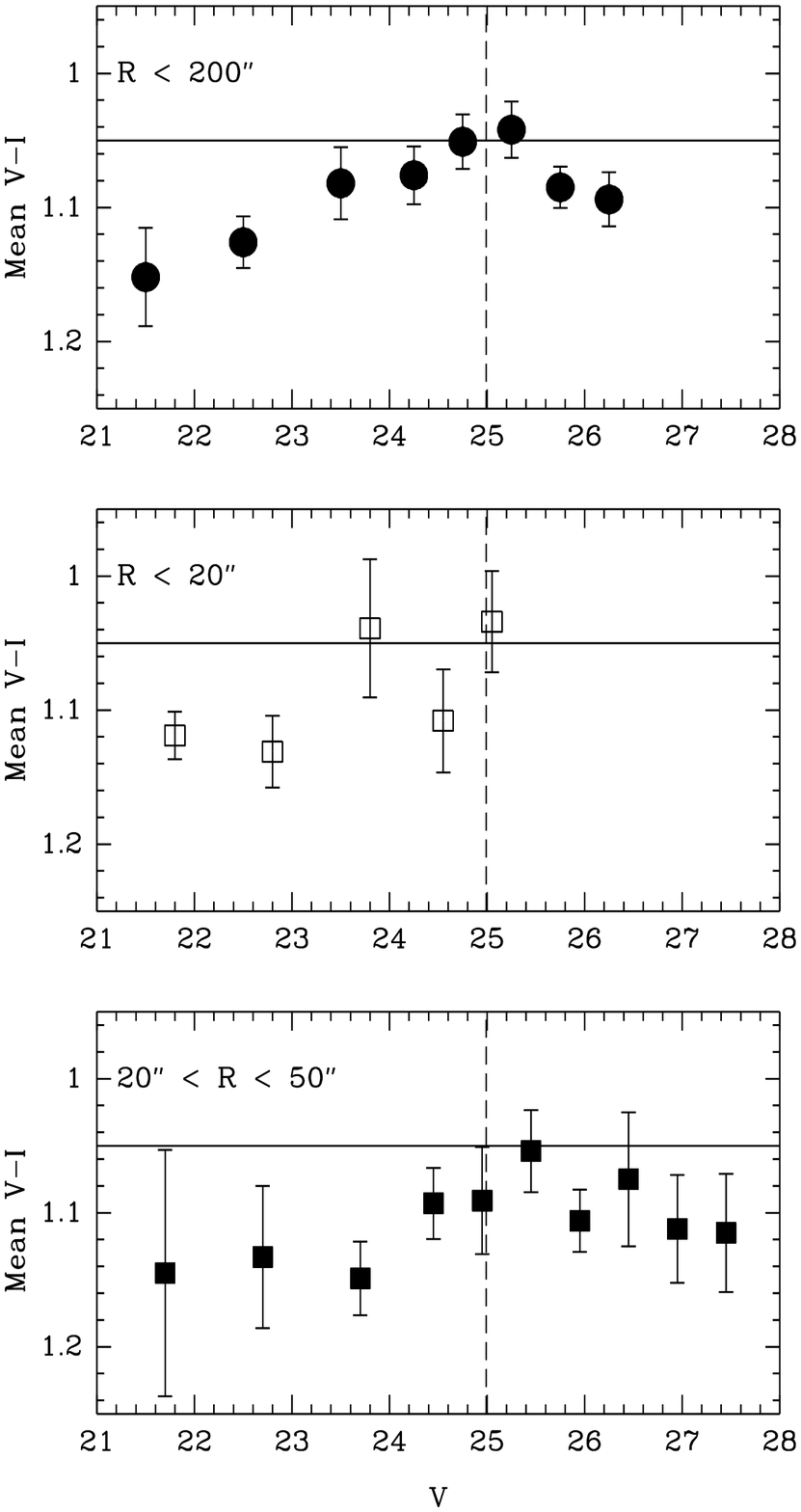}
\vspace{18.0cm}
\caption{}
\end{figure*}

\clearpage
\begin{figure*}
\includegraphics{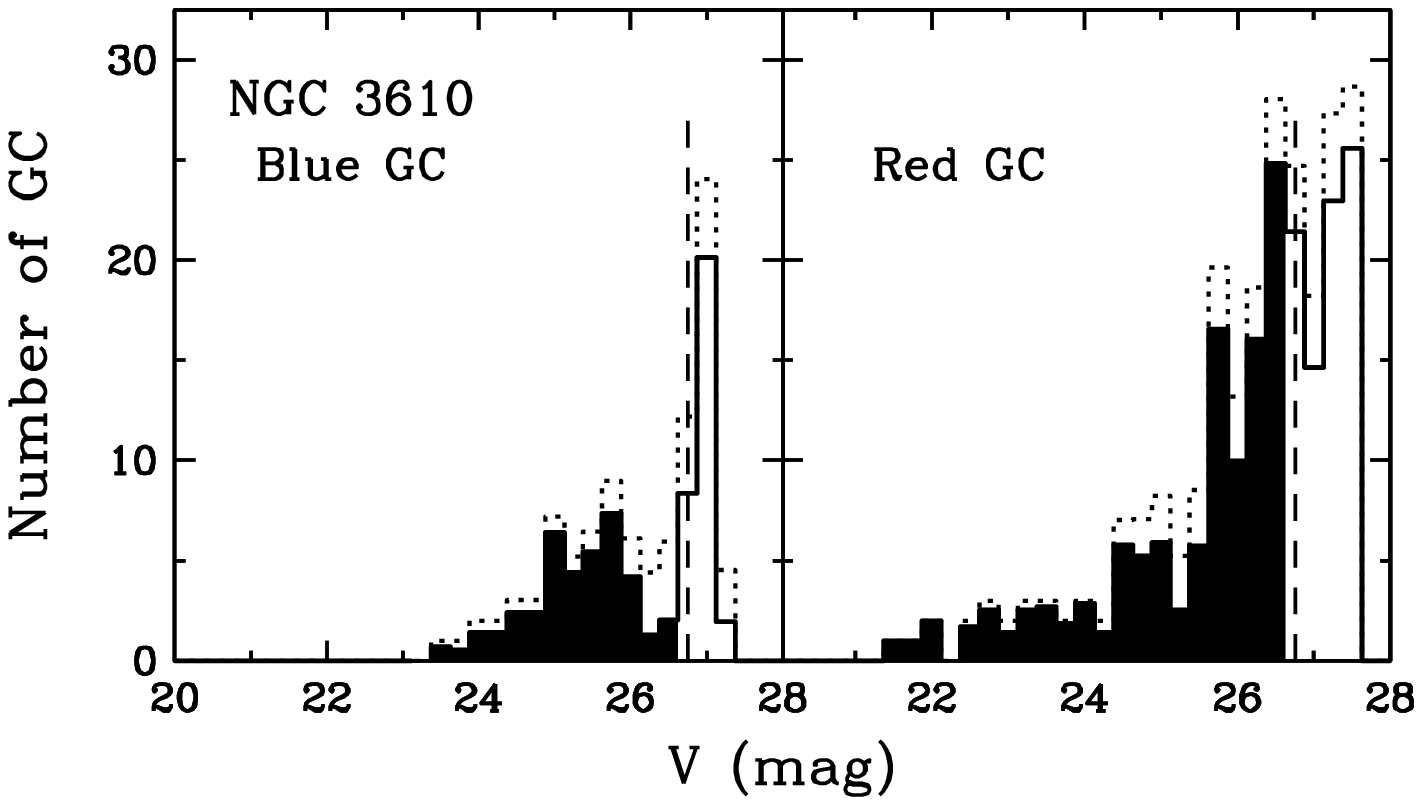}
\vspace{17.0cm}
\caption{}
\end{figure*}

\clearpage
\begin{figure*}
\includegraphics{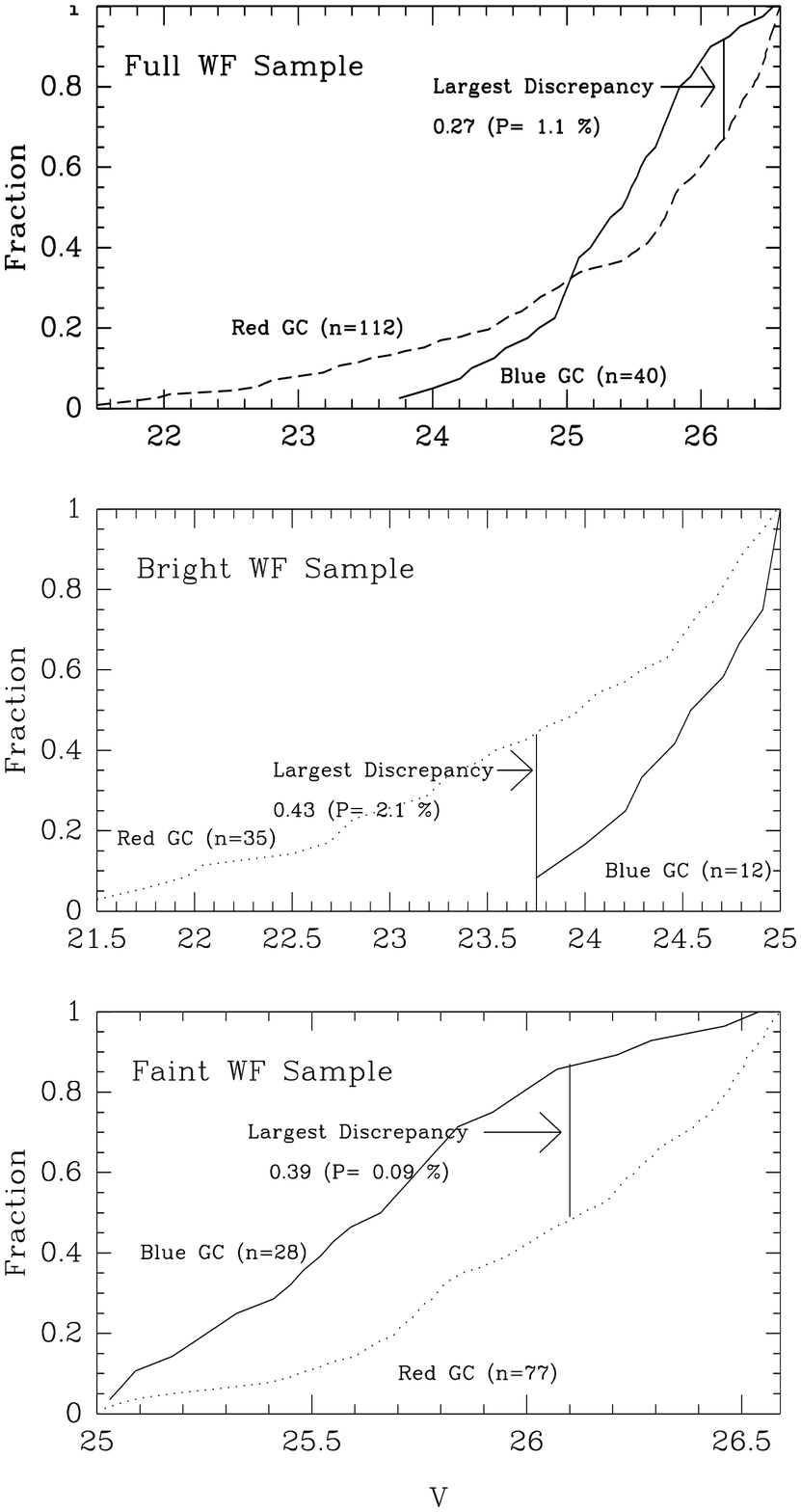}
\vspace{18.0cm}
\caption{}
\end{figure*}

\clearpage
\begin{figure*}
\includegraphics{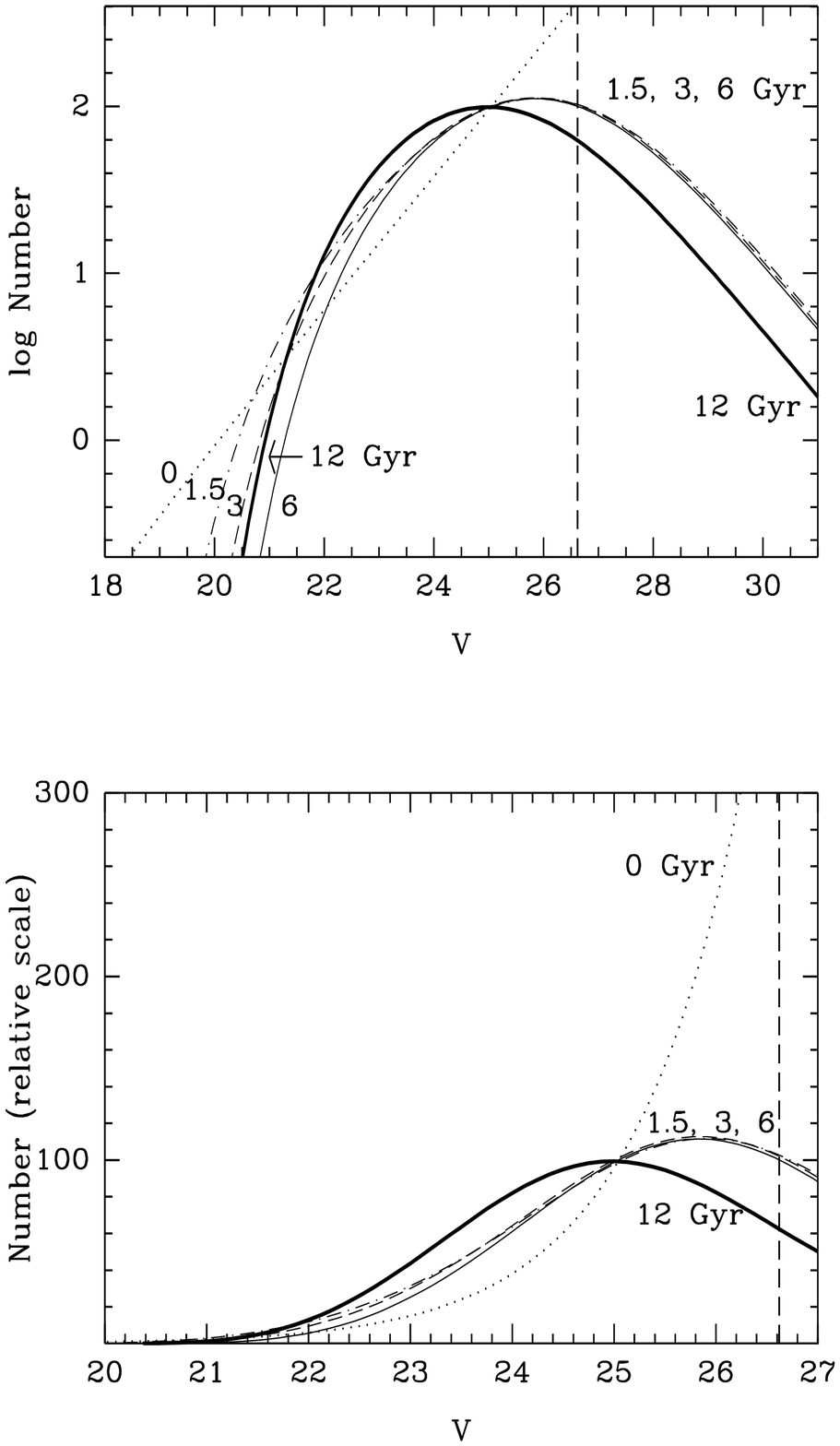}
\vspace{17.0cm}
\caption{}
\end{figure*}

\clearpage
\begin{figure*}
\includegraphics{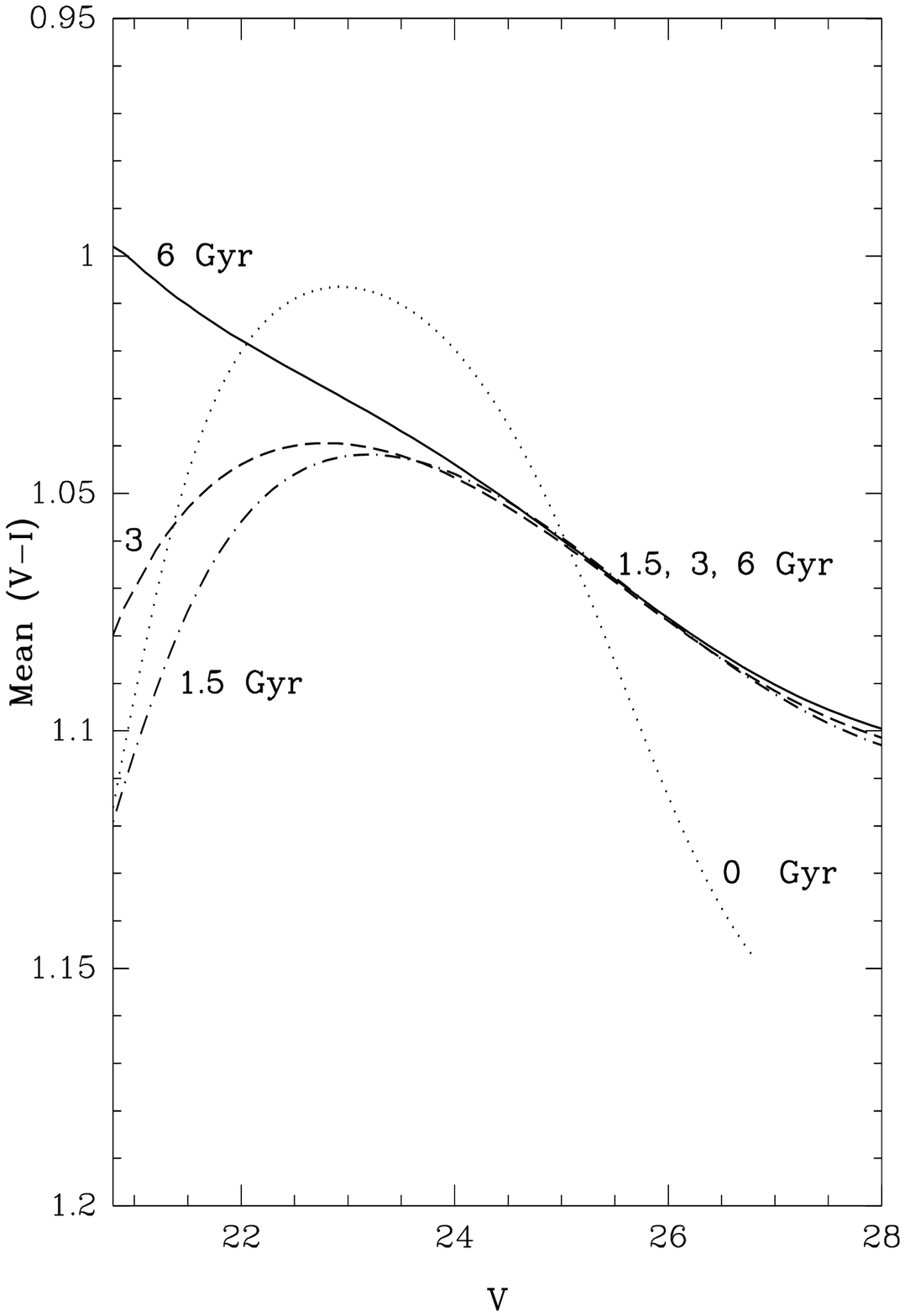}
\vspace{17.0cm}
\caption{}
\end{figure*}

\clearpage
\begin{figure*}
\includegraphics{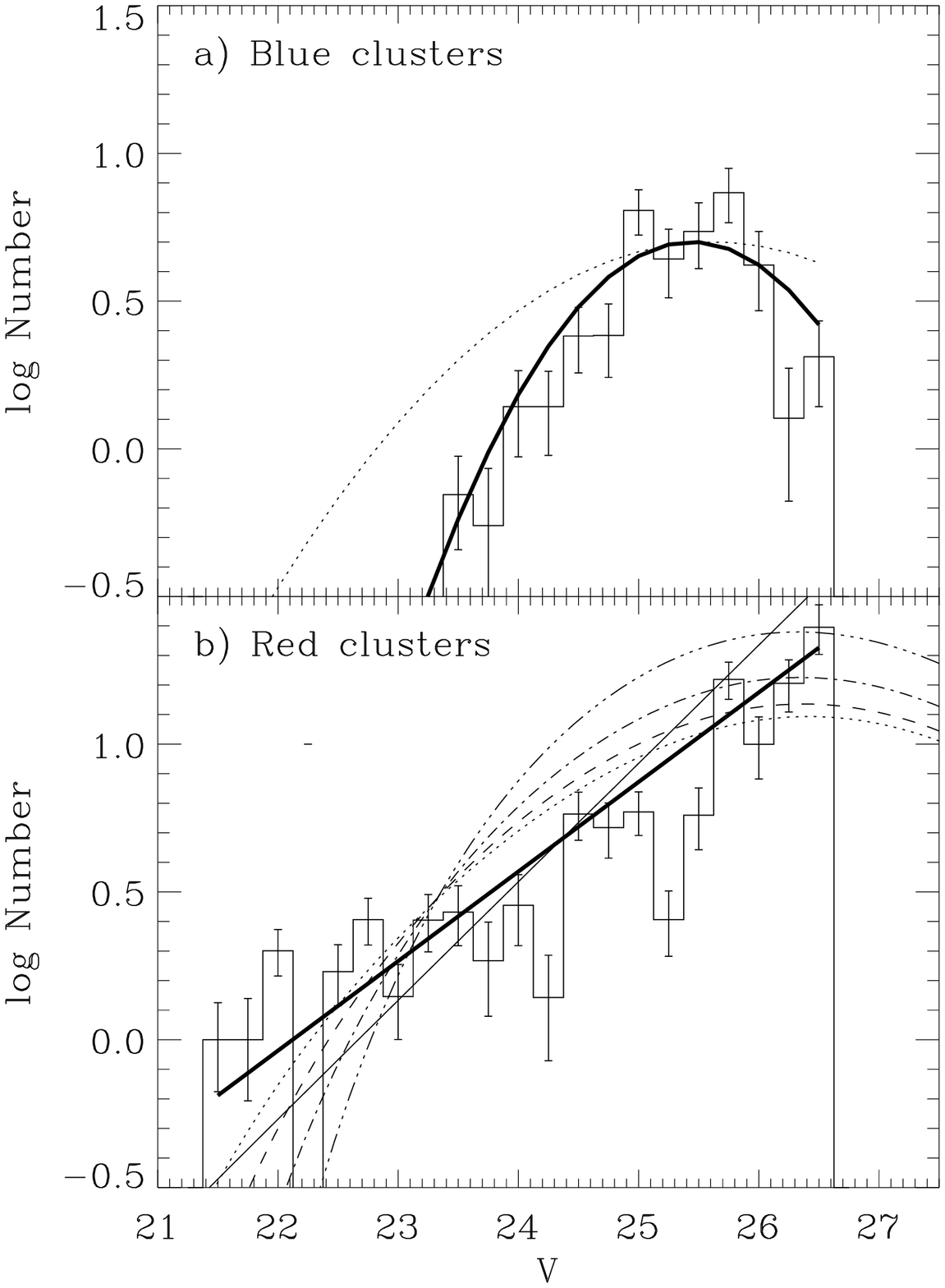}
\vspace{17.0cm}
\caption{}
\end{figure*}

\clearpage
\begin{figure*}
\includegraphics{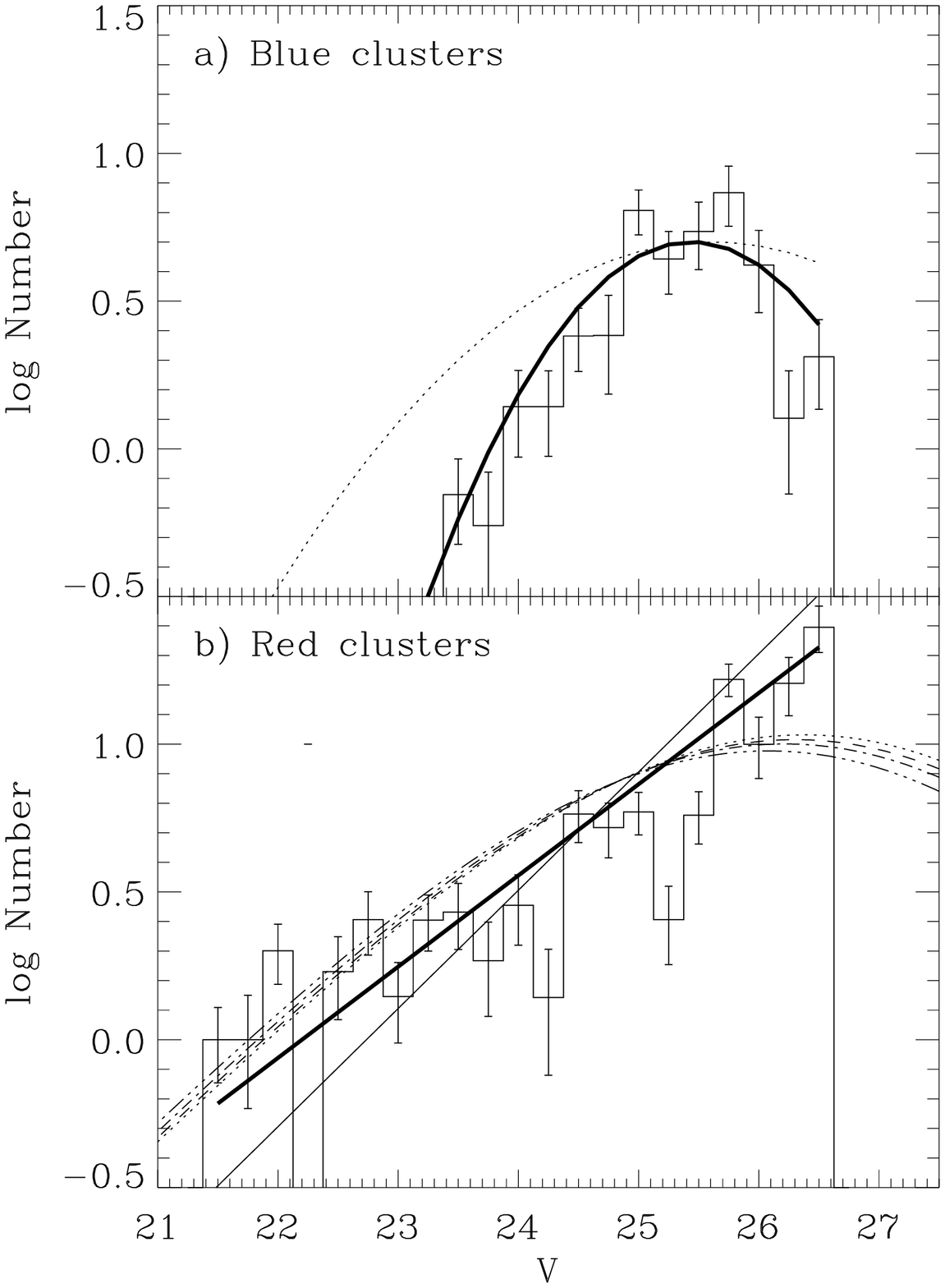}
\vspace{17.0cm}
\caption{}
\end{figure*}

\clearpage
\begin{figure*}
\includegraphics{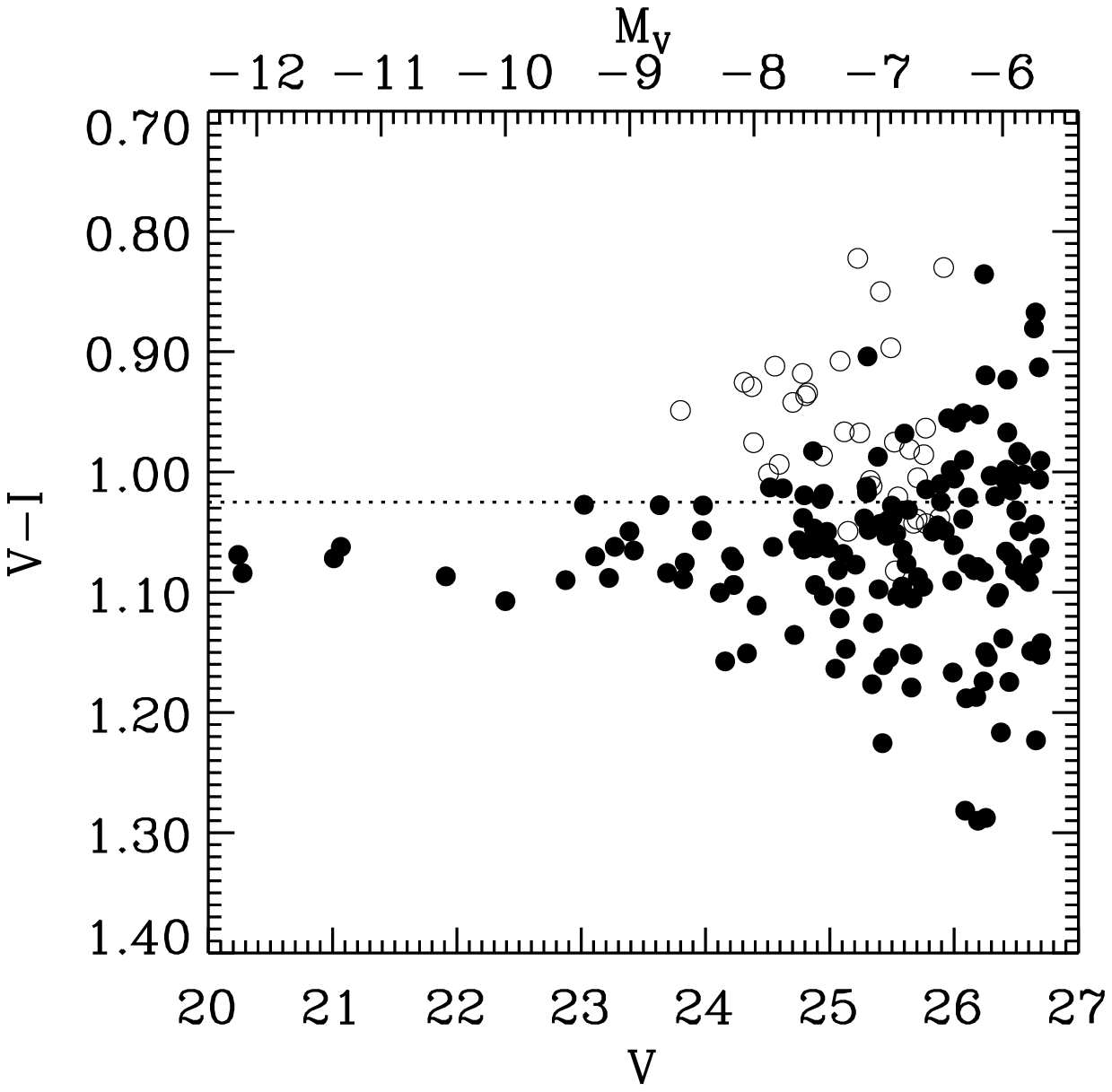}
\vspace{17.0cm}
\caption{}
\end{figure*}

\clearpage
\begin{figure*}
\includegraphics{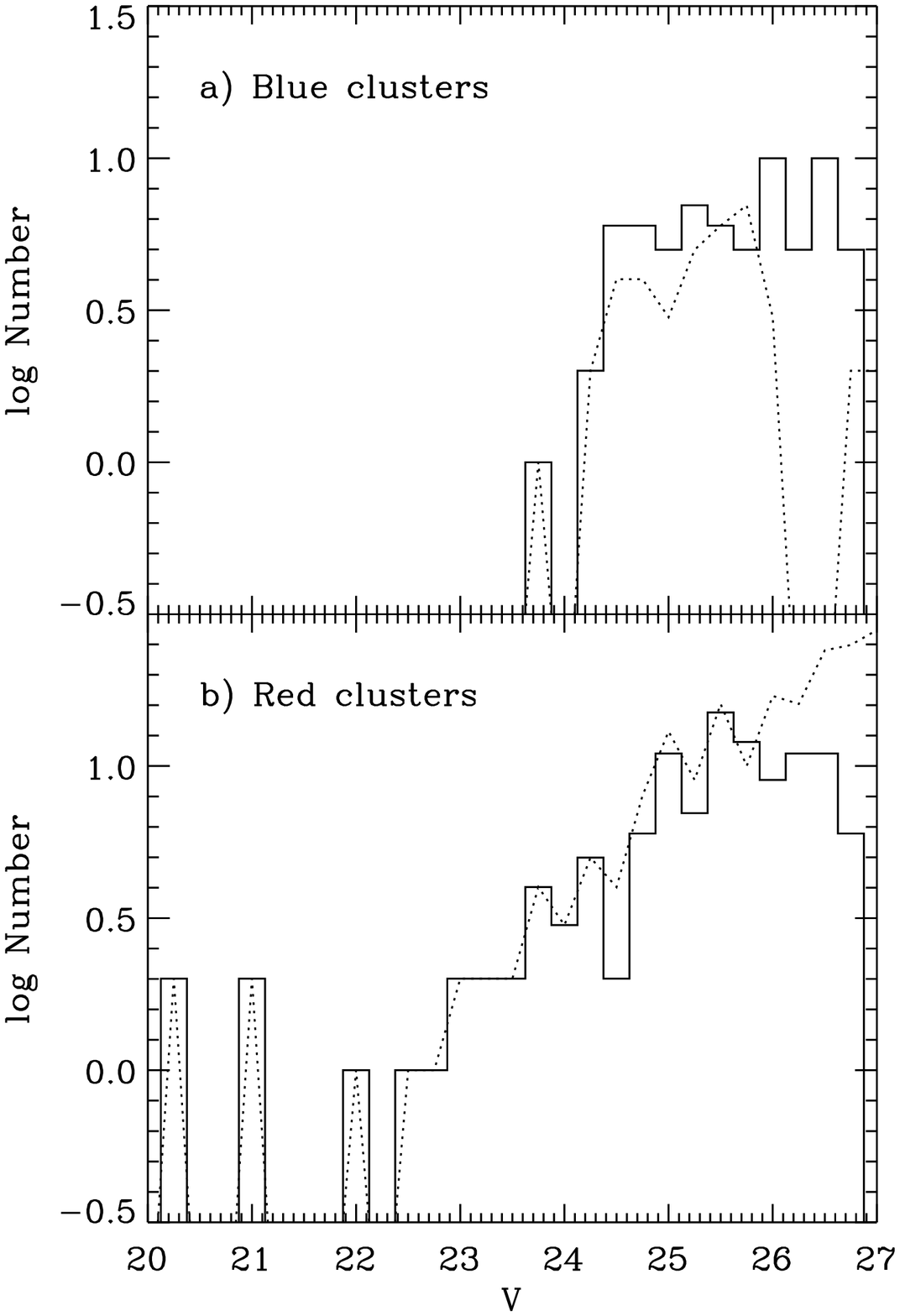}
\vspace{17.0cm}
\caption{}
\end{figure*}

\clearpage 
\begin{figure*}
\includegraphics{whitmore.tab1.ps}
\vspace{17.0cm}
\end{figure*}

\end{large}			

\end{document}